\documentclass{jfm}
\usepackage{graphicx}
\usepackage{epstopdf, epsfig}

\usepackage[normalem]{ulem}
\usepackage[shortlabels]{enumitem}
\usepackage{CJK}
\usepackage{amssymb,euscript,latexsym,amsmath}
\usepackage{color}
\usepackage{epstopdf}
\usepackage{setspace}
\usepackage{chngpage}

\usepackage{chngcntr}

\DeclareMathOperator*{\argmin}{arg\,min}
\DeclareMathOperator*{\argmax}{arg\,max}

\usepackage[usenames,dvipsnames]{xcolor}

\shorttitle{Optimal slip velocities of micro-swimmers with arbitrary axisymmetric shapes}
\shortauthor{H. Guo, H. Zhu, R. Liu, M. Bonnet and S. Veerapaneni}

\title{Optimal slip velocities of micro-swimmers with arbitrary axisymmetric shapes}
\author{Hanliang Guo\aff{1}, 
Hai Zhu\aff{1}, 
Ruowen Liu\aff{1}, 
Marc Bonnet\aff{2}, 
\and Shravan Veerapaneni\aff{1}\corresp{\email{shravan@umich.edu}}}
\affiliation{\aff{1} Department of Mathematics, University of Michigan, Ann Arbor, MI, 48109 USA.
\aff{2}POEMS (CNRS, INRIA, ENSTA), ENSTA Paris, 91120 Palaiseau, France.
}

\begin{document}

\maketitle
%\date{\today}

\begin{abstract}
This article presents a computational approach for determining the optimal slip velocities on any given shape of an axisymmetric micro-swimmer suspended in a viscous fluid. The objective is to minimize the power loss to maintain a target swimming speed, or equivalently to maximize the efficiency of the micro-swimmer. Owing to the linearity of the Stokes equations governing the fluid motion, we show that this PDE-constrained optimization problem reduces to a simpler quadratic optimization problem, whose solution is found using a high-order accurate boundary integral method. We consider various families of shapes parameterized by the reduced volume and compute their swimming efficiency. {Among those, prolate spheroids were found to be the most efficient micro-swimmer shapes for a given reduced volume. We propose a simple shape-based scalar metric that can determine whether the optimal slip on a given shape makes it a pusher, a puller or a neutral swimmer.} 
\end{abstract}

\section{Introduction}
\label{sc:intro}

The {\em squirmer model} \citep{Lighthill1952Squirming, Blake1971Spherical} 
is widely adopted by mathematicians and physicists over the past decades to model ciliated micro-swimmers such as \textit{Opalina}, \textit{Volvox} and \textit{Paramecium}~\citep{Lauga2009hydrodynamics}. On a high level, this continuum model, sometimes referred to as the {\em envelope model}, effectively tracks the motion of the envelope formed by the tips of the densely-packed cilia, located on the swimmer body, while neglecting the motion below the tips. Individual and collective ciliary motions could be mapped to traveling waves of the envelope on the surface.  
Assuming no radial displacements of the surface and time-independent tangential velocity led to the simpler {\em steady} squirmer model \citep[see][]{Pedley2016Spherical}, wherein, a prescribed slip velocity on the boundary propels the squirmer. While the model was originally designed for spherical shapes, it has since been adapted to more general shapes and has recently been shown to capture realistic collective behavior of suspensions \citep{Kyoya2015Shape}.  

Shape is also a key parameter in the design of artificial micro-swimmers for {promising }applications such as targeted drug delivery. In particular, the squirmer model is often employed to study the propulsion of {\em phoretic particles}, which are micro- to nano-meter sized particles that propel themselves by exploiting the asymmetry of chemical reactions on their surfaces~\citep{Anderson1989Colloid, Golestanian2007Designing}. A classical example is the Janus sphere~\citep{Howse2007Self}, which consists of inert and catalytic hemispheres. When submerged in a suitable chemical solution, the asymmetry between the chemical reactions on the two hemispheres creates a concentration gradient. The gradient creates an effective steady slip velocity on the surface via osmosis that naturally suits the squirmer model. Besides the classical Janus spheres and bi-metallic nanorods~\citep{Paxton2004Catalytic}, more sophisticated shapes have also been proposed recently, such as two-spheres~\citep{Valadares2010Catalytic, Palacci2015Artificial}, spherocylinder~\citep{Uspal2018Shape}, matchsticks~\citep{Morgan2014Chemotaxis} and microstars~\citep{Simmchen2017Dynamics}. 
Interestingly, \citet{Uspal2018Shape} showed that special shapes of phoretic particles exhibit novel properties such as `edge-following' when put close to chemically patterned surfaces.

Studying the efficiency of biological micro-swimmers is pivotal to understanding natural systems and designing artificial ones for accomplishing various physical tasks.
The mechanical efficiency~\citep{Lighthill1952Squirming} of the spherical squirmer can be directly computed, as its rate of viscous energy dissipation, or power loss, can be written in terms of the modes of the squirming motion.
\citet{Michelin2010Efficiency} found the optimal swimming strokes of unsteady spherical squirmers by employing a pseudo-spectral method for solving the Stokes equations that govern the ambient fluid and 
a numerical optimization procedure. Their approach, however, does not readily generalize to arbitrary shapes. On the other hand, \citet{leshansky2007frictionless} analytically investigated the efficiency of micro-swimmers of prolate {spheroids} shapes with a time-independent  `treadmilling' slip velocity and found that the efficiency increases unboundedly with the aspect ratio.
\citet{Vilfan2012Optimal} optimized the steady slip velocity and the shape at the same time, with constraints on its volume and maximum curvature. 
{The work considered power loss not only outside but also inside the squirmer surface, which could be an order of magnitude higher than the outside power loss alone~\citep{keller1977porous, ito2019swimming}. However, it assumed that the tangential force on the squirmer surface is linear to its {\em local} slip velocity, which is not always the case for microswimmers.}

In this paper, we address the following broader questions:
{\em Given an axisymmetric shape of a steady squirmer, what is the slip velocity that maximizes its swimming efficiency? 
}
The optimization problem, being quadratic, is reduced to a linear system of equations solved by a direct method, while forward exterior flow problems are solved using a boundary integral method. Those combined features produce a simple and efficient solution procedure.
We introduce the optimization problem and our numerical solver in Section \ref{sc:problem}, present the optimal solution for various shape families, {summarize the correlations between the shapes and the optimal slip velocities, and propose a shape-based scalar metric to predict whether the optimized swimmer would be a pusher or a puller} in Section \ref{sc:results}, followed by conclusions and a discussion on future research directions in Section \ref{sc:future}.

\section{Problem Formulation and Numerical Solution} \label{sc:problem}
\subsection{Model} 
\begin{figure}
        \centerline{\includegraphics[width=.8\linewidth]{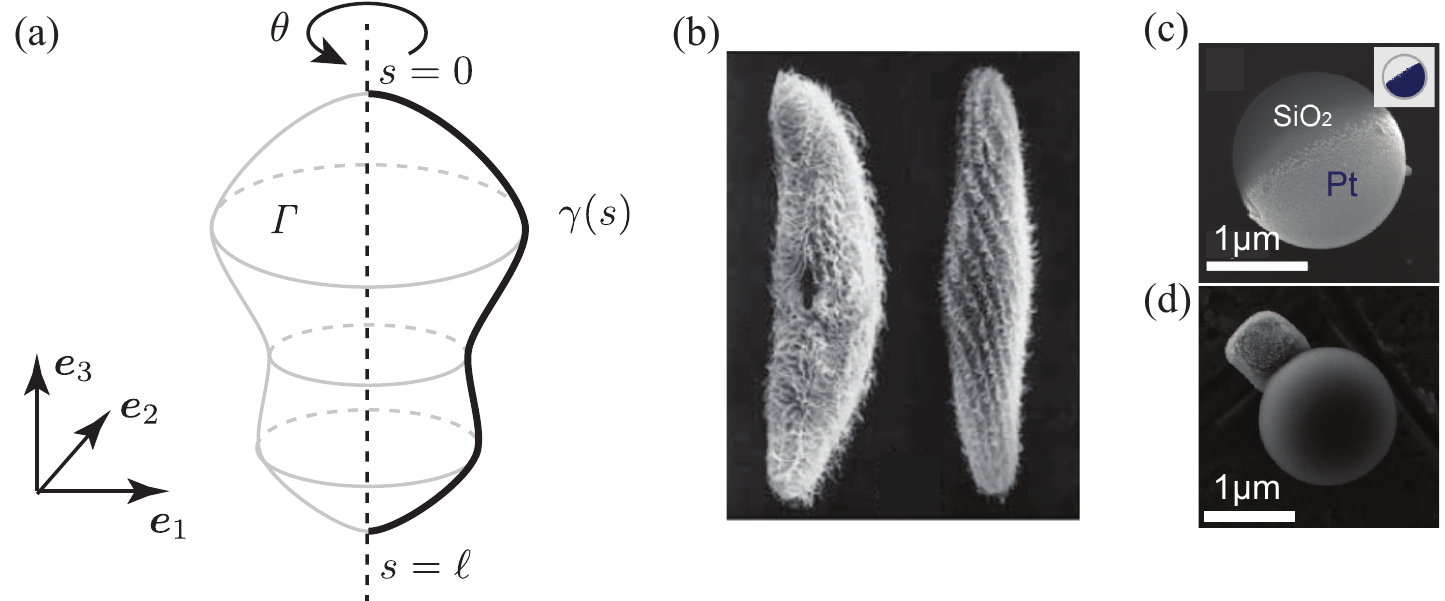}}
\caption[]{(a) Schematic of the micro-swimmer geometry. The shape is assumed to be axisymmetric, obtained by rotating the generating curve $\gamma$ about the $\boldsymbol{e}_3$ axis. 
	(b)  Biological swimmers (\citet{lynn2008ciliated}, Chap 4 Fig 4.6).
	(c) Scanning electron microscope (SEM) image of a single half-coated Janus particle; inset: the dark-blue shows the location of the Pt cap. \citep{Choudhury2017Active}
	(d) SEM image of a phototactic swimmer, which consists of a haematite particle extruded from a colloidal bead. \citep{aubret2018diffusiophoretic}}
	\label{fig:model}
\end{figure}

Consider an axisymmetric micro-swimmer whose boundary $\Gamma$ can be obtained by rotating a curve $\gamma$ about $\boldsymbol{e}_3$ axis 
 as shown in Fig.~\ref{fig:model}(a). Using the arc-length $s\in[0,\ell]$ to parameterize the generating curve, its coordinate functions can be written as $\gamma(s) = (x_1(s), 0, x_3(s))$. Here, we restrict our attention to shapes of spherical topology, therefore, all shapes considered satisfy the conditions $x_1(0) = x_1(\ell) = 0$ and $x_1(s)>0, \,\, \forall \, s\in(0,\ell)$. We assume that the micro-swimmer is suspended in an unbounded viscous fluid domain. The governing equations for the ambient fluid in the vanishing
Reynolds number limit are given by the Stokes equations:
\begin{equation}\label{eq:stokes}
-\mu\nabla^2\boldsymbol{u} + \nabla p = \boldsymbol{0},\quad \nabla\cdot\boldsymbol{u} = 0,
\end{equation}
where $\mu$ is the fluid viscosity, $p$ and $\boldsymbol{u}$ are the pressure and flow field respectively.  In the absence of external forces and imposed flow fields, the far-field boundary condition simply is 
\begin{equation} \label{eq:far-bc} 
\lim_{\boldsymbol{x}\rightarrow\infty}\boldsymbol{u}(\boldsymbol{x}) = \boldsymbol0. \end{equation}
%---------------------
%------------------------
A tangential slip ${u}^{\mathrm{S}}$ defined on $\gamma$ propels the micro-swimmer forward with a translational velocity $U$ in the $\boldsymbol{e}_3$ direction. Its angular velocity as well as the translational velocities in the $\boldsymbol{e}_1$ and $\boldsymbol{e}_2$ directions are zero by symmetry. Consequently, the boundary condition on $\gamma$ is given by 
%--------------
\begin{equation}\label{eq:bc}
\boldsymbol{u} = {u}^{\mathrm{S}}\boldsymbol\tau+{U}\boldsymbol{e}_3,
\end{equation}
%--------------
where $\boldsymbol\tau$ is the unit tangent vector on $\gamma$.  Note that, in order to avoid singularities, the slip must vanish at the end points:
\begin{equation}\label{eq:slip}
u^{\mathrm{S}}(0) = u^{\mathrm{S}}(\ell) = 0.
\end{equation}

Due to the axisymmetry of $\Gamma$, the required no-net-torque condition on the freely-suspended micro-swimmer is automatically satisfied while the no-net-force condition reduces to one scalar equation
\begin{equation}\label{eq:nonetforce}
\int_\Gamma \boldsymbol{f}(\boldsymbol{x})\cdot\boldsymbol{e}_3 \, \mathrm{d}S=2\pi\int_\gamma {f}_3(\boldsymbol{x})\, x_1\mathrm{d}s= 0,
\end{equation}
where $\boldsymbol{f}$ is the active force density on the micro-swimmer surface (negative to fluid traction) and $f_3$ is its $\boldsymbol{e}_3$ component.

%%% power loss
We quantify the performance of the micro-swimmer with slip velocity $u^{\mathrm{S}}$ by its power loss while maintaining a target swimming speed $U$.
The power loss is defined by
\begin{equation}\label{eq:P}
P = \int_\Gamma\boldsymbol{f}\cdot\boldsymbol{u} \,\mathrm{d}S
= 2\pi\int_\gamma \boldsymbol{f}\cdot({u}^{\mathrm{S}}\boldsymbol\tau+U\boldsymbol{e}_3) x_1\mathrm{d}s.
\end{equation}
Note that $P$ can be made arbitrarily small by lowering the swimming speed $U$. It is therefore necessary to compare the power loss of different swimmers that have the same swimming speed $U$.
We note that a lower $P$ with a fixed shape and swimming speed $U$ corresponds to a higher efficiency, $\eta = C_DU^2/P$, as defined by~\citet{Lighthill1952Squirming}, where $C_D$ is the drag coefficient of the given swimmer.

\subsection{Boundary integral method for the forward problem}
Before stating the optimization problem, we summarize our numerical solution procedure for (\ref{eq:stokes}) -- (\ref{eq:bc}). 
Again, we fix the swimming speed $U$, referred to from here onwards as the``target swimming speed'', and assume that the tangential slip $u^S$ is given. In general, an arbitrary pair of ${u}^{\mathrm{S}}$ and $U$ does not satisfy the no-net-force condition~\eqref{eq:nonetforce}. This condition will be treated as a constraint in our optimization problem. Therefore, 
the goal is to find the active force density $\boldsymbol{f}$ given the velocity on the boundary $\gamma$ as in (\ref{eq:bc}). We use the single-layer potential ansatz, which expresses the velocity as a convolution of an unknown density function $\boldsymbol{\mu}$ with the Green's function for the Stokes equations {$G$}, from which the force density can be determined by convolution with the traction kernel {$T$}:
\begin{equation} \boldsymbol{u}(\boldsymbol{x}) = \int_\Gamma G(\boldsymbol{x} - \boldsymbol{y}) \, \boldsymbol{\mu}(\boldsymbol{y}) \, d\Gamma(\boldsymbol{y}), \quad  \boldsymbol{f}(\boldsymbol{x}) 
=-\frac{1}{2}\boldsymbol{\mu}\left(\boldsymbol{x}\right)+\boldsymbol{n}\left(\boldsymbol{x}\right)\int_{\Gamma}T\left(\boldsymbol{x}-\boldsymbol{y}\right)\boldsymbol{\mu}\left(\boldsymbol{y}\right)d\Gamma\left(\boldsymbol{y}\right)
,\label{eq:BI}\end{equation}
where $\boldsymbol{n}$ is the unit normal vector pointing into the fluid.
We can solve for $\boldsymbol{\mu}$ by taking the limit of $\boldsymbol{x} \rightarrow \Gamma$ in the above ansatz and substituting in (\ref{eq:bc}). The boundary integrals in (\ref{eq:BI}) become weakly singular on $\Gamma$, requiring specialized quadrature rules. Here, we use the approach of \citet{veerapaneni2009numerical} which performs an analytic integration in the $\theta-$direction reducing the integrals to convolutions on the generating curve and applies a high-order quadrature rule designed to handle the $log-$singularity of the resulting kernels.  {More details on the numerical scheme are provided in Appendix~\ref{sec:validation}.}

\subsection{Optimization problem and its reformulation}
\label{slip:quadratic}
The goal is to find a slip profile $u^{\mathrm{S}*}(s)$ that minimizes the power loss $P$ while maintaining the target swimming speed $U$ of a given axisymmetrical micro-swimmer.
Let ${J}$ be the objective function, here equated to $P$ defined in (\ref{eq:P}), and ${F}$ be the net force functional:
\begin{equation}\label{eq:powerloss}
{J}(u^{\mathrm{S}}) := 2\pi\int_\gamma \boldsymbol{f}(u^{\mathrm{S}})\cdot(u^{\mathrm{S}}\boldsymbol\tau + U\boldsymbol{e}_3)\, x_1 \mathrm{d}s, 
\quad 
{F}(u^{\mathrm{S}}) :=2\pi \int_\gamma \boldsymbol{f}(u^{\mathrm{S}}) \cdot \boldsymbol{e}_3\, x_1 \mathrm{d}s.
\end{equation}
They are slip velocity functionals as their values are completely determined by $u^{\mathrm{S}}$.  The optimization problem can now be stated as follows: 
\begin{equation}\label{eq:opt}
u^{\mathrm{S*}}=\argmin_{u^{\mathrm{S}}\in \mathcal{U}} {J}(u^{\mathrm{S}})\quad \text{ subject to } {F}(u^{\mathrm{S}}) = 0,
\end{equation}
with $\mathcal{U}$ being the space of the all possible slip velocities satisfying \eqref{eq:slip}.  Notice that the no-net-force condition (\ref{eq:nonetforce}) is added as a constraint here. 

By \eqref{eq:bc} and linearity of the Stokes equation~\eqref{eq:stokes}, the forward solution $\boldsymbol{u}$ and the net force ${F}$ are affine in $u^{\mathrm{S}}$ ($\boldsymbol{u}$ is linear in $u^{\mathrm{S}}$ if $F=0$). Consequently, ${J}(u^{\mathrm{S}})$ is a quadratic functional and (\ref{eq:opt}) is inherently a quadratic optimization problem. To make it more explicit,  consider a discretized version of the slip optimization problem where $u^{\mathrm{S}}$ is sought in the form
\begin{equation}\label{eq:disc_slip}
  u^{\mathrm{S}}(\boldsymbol{x}) = \sum_{k=1}^{m} U\xi_k\, u^{\mathrm{S}}_k(s),
\end{equation}
for some set of $m$ {basis functions} $u^{\mathrm{S}}_k$ satisfying \eqref{eq:slip}. We adopt a B-spline formulation for these basis functions {(see Appendix~\ref{sec:parameterspace} for more details)}.
Let $(\boldsymbol{u}_0,p_0,\boldsymbol{f}_0)$ and $(\boldsymbol{u}_k,p_k,\boldsymbol{f}_k)$ (with $1\le k\le m$) denote the solutions of the forward problem~\eqref{eq:stokes} with $\boldsymbol{u}=\boldsymbol{e}_3$ and $\boldsymbol{u}=u_k^{\mathrm{S}}\boldsymbol{\tau}$ being their boundary conditions on $\gamma$, respectively. 

The net force ${F}(u^{\mathrm{S}})$ is then given by  ${F}(u^{\mathrm{S}})=  2\pi U\mathcal{F}(\boldsymbol\xi)$, where
\begin{equation}
 \mathcal{F}(\boldsymbol\xi):= \int_\gamma \left(\boldsymbol{f}_0 + \sum_{k = 1}^m \xi_k\boldsymbol{f}_k \right)\cdot \boldsymbol{e}_3\, x_1 \mathrm{d}s= F_0 + \boldsymbol{F}^{\mathrm{T}}\boldsymbol{\xi}.
\end{equation}
Here $\boldsymbol{\xi}=(\xi_1,\ldots,\xi_m)^{\mathrm{T}}$, $\boldsymbol{F} = (F_1,\ldots,F_m)^{\mathrm{T}}$, and $F_k = \int_\gamma \boldsymbol{f}_k \cdot \boldsymbol{e}_3\, x_1 \mathrm{d}s$ for $k=0,1,\cdots,m$.

Similarly, we have ${J}(u^{\mathrm{S}}) =  2\pi U^2\mathcal{J}(\boldsymbol\xi)$, where
\begin{equation}
\mathcal{J}(\boldsymbol\xi):= \int_\gamma \left(\boldsymbol{f}_0 + \sum_{k = 1}^m \xi_k\boldsymbol{f}_k \right)\cdot\left(\boldsymbol{e}_3 + \sum_{j=1}^m \xi_j u^{\mathrm{S}}_j\boldsymbol\tau\right)\, x_1 \mathrm{d}s = \boldsymbol{\xi}^{\mathrm{T}}\boldsymbol{A}\boldsymbol{\xi} + 2\boldsymbol{\xi}^{\mathrm{T}}\boldsymbol{F} + F_0.
\end{equation}
The elements of the $m\times m$ matrix $\boldsymbol{A}$ are given by $A_{kj} = \int_\gamma \boldsymbol{f}_k \cdot u^{\mathrm{S}}_j \boldsymbol\tau\, x_1 \mathrm{d}s$.
We have used the fact that $\int_\gamma \boldsymbol{f}_0 \cdot u^{\mathrm{S}}_k \boldsymbol\tau\, x_1 \mathrm{d}s = \int_\gamma \boldsymbol{f}_k \cdot \boldsymbol{e}_3\, x_1 \mathrm{d}s$ for the linear term by the reciprocal theorem~\citep{happel1973low}.
We note that $\boldsymbol{A}$ is symmetric, also by the reciprocal theorem.
{Physically speaking, $\boldsymbol\xi^\mathrm{T}\boldsymbol{A}\boldsymbol\xi$ represents the scaled power loss of the swimmer being held still with its slip velocity parametrized by $\boldsymbol\xi$, implying that $\boldsymbol{A}$ is positive-definite; {$\boldsymbol{\xi}^T \boldsymbol{F}$ is the scaled power loss of the active force along the swimming direction; $F_0$ is the scaled power loss of tolling a rigid body with the same shape as the micro-swimmer at unit speed. }}  

Now, the discretized optimization problem becomes
\begin{equation}
  \min_{\boldsymbol{\xi}\in\mathbb{R}^m} \mathcal{J}(\boldsymbol{\xi}) \quad \text{ subject to } \mathcal{F}(\boldsymbol{\xi})=0. \label{opt:discr}
\end{equation}
Introducing the Lagrangian $L(\boldsymbol{\xi},\lambda):= \mathcal{J}(\boldsymbol{\xi})-2\lambda \mathcal{F}(\boldsymbol{\xi})$, the slip optimization problem is reduced to solving the first-order stationarity equations for $L$ given by
\begin{equation}
  \begin{bmatrix} \boldsymbol{A} & -\boldsymbol{F} \\ -\boldsymbol{F}^{\mathrm{T}} & 0 \end{bmatrix} \, \begin{bmatrix} \boldsymbol{\xi} \\ \lambda \end{bmatrix}
 = \begin{bmatrix} -\boldsymbol{F} \\ F_0 \end{bmatrix}. \label{stat:eqn}
\end{equation}
Note that forming the matrix requires $(m+1)$ solves of the forward problem (\ref{eq:stokes}) with appropriate boundary conditions. Since the micro-swimmer is assumed to be rigid, the single layer potential operator as well as the traction operator, required for forming $\boldsymbol{A}$ and $\boldsymbol{F}$, are both fixed for a given shape. Therefore, we only need to form them once. 
 
%%%%%%%%%%%%%%%%%%%%%%%%%%%

\section{Results}
\label{sc:results}
\begin{figure}
        \centerline{\includegraphics[scale = 1]{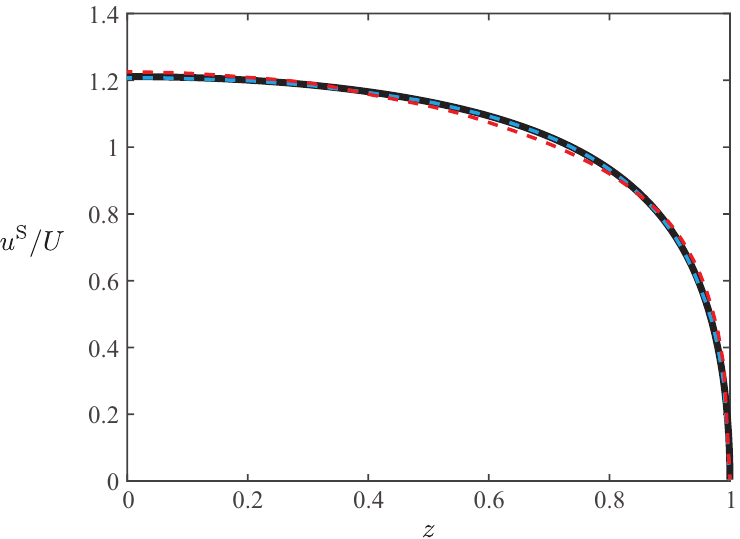}}
\caption[]{Optimal slip velocity compared to \citet[Figure 4]{leshansky2007frictionless}. The aspect ratio of the prolate spheroid is $(1+2.5^2)^{1/2}$. Our numerical optimization is depicted in black solid curve, while dash curves represent analytical solutions at different truncation levels $L=4$ (red) and $L=10$ (blue).}
	\label{fig:leshansky}
\end{figure}

{We tested the convergence of our numerical solvers rigorously}; the boundary discretization for all the numerical examples presented here 
is chosen so that at least 6-digit solution accuracy is attained (determined {\em via} self-convergence tests). 
{The optimal slip velocity for a particular prolate spheroid tested against the (truncated) analytical solution given by \citet{leshansky2007frictionless} is shown in Fig.~\ref{fig:leshansky}. Our numerical solution is indistinguishable against the analytical solution at their finer truncation level $L=10$.}
{Additional validation results can be found in the Appendix~\ref{sec:validation}}.

Here we focus on analysis of the optimal solutions for various micro-swimmer shape families. Let $V$ be the volume enclosed by the swimmer. We normalize lengths by the radius of a sphere of equivalent volume i.e., by $R=(3V/4\pi)^{1/3}$, and velocities by the swimming speed $U$. A simple calculation shows that, for a micro-swimmer submerged in water of size $R=5\,\mu\text{m}$ and the speed of one body-length per second, the Reynolds number (Re) $ \approx 5\times 10^{-5}$; thereby, confirming the validity of the Stokes equation~\eqref{eq:stokes}. We will use the dimensionless {\em reduced volume}, defined by $\nu = 6\sqrt{\pi}V/A^{3/2}$ where $A$ is the surface area of the given shape, to characterize each shape family. The largest possible value of $\nu$, attained by spheres, is $\nu=1$, while for example $\nu$ decreases monotonically for {spheroids} as the aspect-ratio is increased.

%---------------------
%------------------------

\begin{figure}
        \centerline{\includegraphics[width=.9\textwidth]{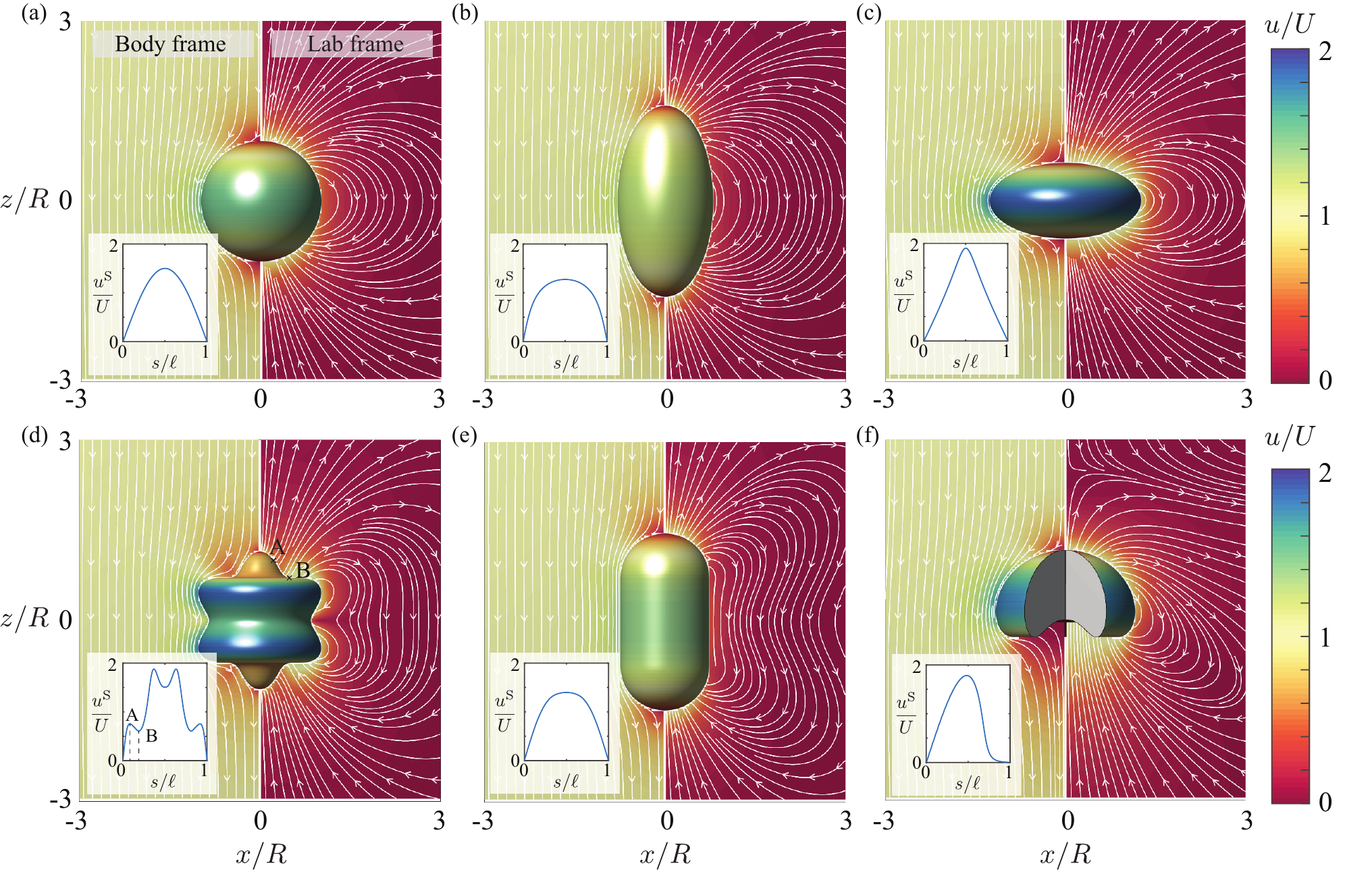}}
	\caption[]{Flow fields and the optimal slip velocity for a few swimmers with typical shapes: (a) Sphere, (b) Prolate {spheroid}, (c) Oblate {spheroid}, (d) Wavy, (e) Spherocylinder, (f) Stomatocyte. Insets show the optimal slip velocities as functions of arc-length along the generating curve. The optimization is performed using 21 control points on the generating curve for representing the slip velocity. {The colormap holds for both the slip velocity and the flow fields.}}
	\label{fig:example}
\end{figure}

We first consider six different micro-swimmer shapes and plot their optimal slip profiles obtained by solving (\ref{stat:eqn}) in Fig.~\ref{fig:example}. In each case, we also show the flow fields in both the body and lab frames. The optimal slip velocities plotted against the arclength, measured from north pole to south pole, are shown in the insets. In the case of a sphere (Fig.~\ref{fig:example}(a)), we recover the standard result that the optimal profile is a sine curve \citep{Michelin2010Efficiency}. The optimal slip velocity of the prolate swimmer, shown in Fig.~\ref{fig:example}(b), `flattens' the sine curve in the middle while that of the oblate swimmer, shown in Fig.~\ref{fig:example}(c), `pinches' the sine curve. Additionally, the peak value of the optimal slip velocity is low for the prolate swimmer, and high for the oblate swimmer, compared to the spherical swimmer.

Next, we consider three shapes corresponding to different shape families.
In Fig.~\ref{fig:example}(d), we consider the `wavy' configuration obtained by adding high-order axisymmetric modes to the spherical shape. The optimal slip velocity follows the general trend for that of (a), while lower slip velocities are observed at the troughs, qualitatively consistent to those obtained in~\citet{Vilfan2012Optimal}. The spherocylinder (Fig.~\ref{fig:example}(e)) resembles closely the prolate {spheroid} of Fig.~\ref{fig:example}(b) with the same aspect ratio, its optimal
slip velocity being
nearly the same (albeit with a slightly narrower plateau and higher peak slip velocity).
Finally, we investigate the optimal slip velocity of the stomatocyte shape (Fig.~\ref{fig:example}(f)), which is the only non-convex shape among those considered here. Similar to that of the oblate swimmer, the general slip velocity is like a pinched sine wave. However, one distinguishing feature is that slip velocity is nearly zero over part of its surface, namely the cup-like region in its posterior. 
%---------------------
%------------------------

{The optimal slip velocity strongly depends on the local geometry of the micro swimmer. 
Generally speaking, the optimal slip velocity is high if the material point is far away from the axis of symmetry.
This could be seen most clearly in the cases of spheroids Fig.~\ref{fig:example}(a)-(c). Specifically, the peak value of the optimal slip velocity is the highest for the oblate spheroid and lowest for the prolate spheroid among the three. 
Intuitively, an object that has a larger radius would endure a higher fluid drag compare to one with a smaller radius when moving in the same speed. Thus extra effort, in the form of slip velocity, would need to be put in to balance the drag.
Additionally, the slip velocity is high when the orientation of the generating curve aligns with the swimming direction (axis of symmetry), and low otherwise. This is understandable as the slip velocity is constructed to be tangential to the generating curve, and a slip velocity perpendicular to the swimming direction generates little swimming velocity at the cost of additional power loss. 
This could be seen most clearly in the wavy shape Fig.~\ref{fig:example}(d). Specifically, comparing the two points A \& B marked in the panel, although point B has a larger radius than point A, the slip velocity of point B is lower because the orientation of the generating curve is almost perpendicular to the swimming direction.
}

Additionally, we note that the optimal slip velocity is proportional to the target swimming speed $U$ due to linearity of the Stokes equations. As a consequence, while the results only showcase micro-swimmers propelling themselves in the positive $\boldsymbol{e}_3$ direction, the optimal solution $u^{\mathrm{S}*}$ for swimming in the opposite direction is merely a change of sign.

{Micro-swimmers can be loosely classified as {\em pushers} that repel fluid from the body along the axis of symmetry, {\em pullers} that draw fluid to the body along the axis of symmetry, or {\em neutral} swimmers that do not repel or draw fluid along the axis of symmetry~\citep{Lauga2009hydrodynamics}. At first sight, the flow fields for all optimal swimmers studied here seem to be neutral swimmers. A closer look into the stresslet tensor $\boldsymbol{S}$, however, reveals a more interesting story. For axisymmetric swimmer whose swimming direction is $\boldsymbol{e}_3$, the stresslet tensor could be simplified to $\boldsymbol{S} = S(\boldsymbol{e}_3\boldsymbol{e}_3 - \boldsymbol{I})$, where $\boldsymbol{I}$ is the identity matrix. The sign of $S$ characterizes whether the swimmer is a pusher ($S<0$) or a puller ($S>0$). 
}

{It is easy to prove by contradiction that the optimal `front-back symmetric' swimmers can not be pushers nor pullers: flipping the swimming direction would make a pusher into a puller of the same shape with an equal (minimal) power loss, contradict to the unique solution guaranteed by the quadratic nature of the problem.
However, the contradiction does not apply for `front-back asymmetric' swimmers as flipping the swimming direction would essentially change the shape of the swimmer. 
In fact, the optimal `front-back asymmetric' swimmers are not always neutral.
For example, the stomatocyte shown in Fig.~\ref{fig:example}(f) is a puller where the stagnation point in the lab frame's flow field is in front of the micro-swimmer.
}

{
Conventionally, pusher and puller particles have been associated with `tail-actuated' swimmers (e.g. spermatozoa) and `head-actuated' swimmers (e.g. \textit{Chlamydomonas reinhardtii}) respectively~\citep{saintillan2015theory}.
It is however not immediately clear whether a micro-swimmer should be a pusher (tail-actuated) or a puller  (head-actuated) to optimize its efficiency when given an arbitrary shape.
Here, capitalizing on our earlier observation on the dependence of local geometry and optimal slip velocity, we propose a shape-based scalar metric $\mathbb{A}$ that can be used to predict whether the optimal swimmer for a given shape is a pusher or puller without the need of optimization.
Simply speaking, $\mathbb{A}$ quantifies the relative `nominal actuation' of the `head' part and the `tail' part of the swimmer based solely on the swimmer shape:
\begin{equation}
\mathbb{A} = \log\left(\frac{\int_{\gamma_h} \boldsymbol{\tau}\cdot\boldsymbol{e}_3\, x_1^2 \mathrm{d}s/\int_{\gamma_h}  x_1 \mathrm{d}s}{\int_{\gamma_t} \boldsymbol{\tau}\cdot\boldsymbol{e}_3\, x_1^2 \mathrm{d}s/\int_{\gamma_t}  x_1 \mathrm{d}s}\right),
\end{equation}
where the generating curve $\gamma$ is divided into two curves $\gamma = \gamma_h \cup \gamma_t$; $\gamma_h$ represents the generating curve of the head part and $\gamma_t$ represents the generating curve of the tail part. 
The numerator and denominator inside the logarithm function are the surface averages of the nominal actuation for the head and tail part respectively. 
The nominal actuation is stronger if the generating curve aligns with the swimming direction better (larger $\boldsymbol\tau\cdot\boldsymbol{e}_3$), or if the material point is farther away from the axis of symmetry (larger $x_1$). 
For front-back symmetric shapes, we naturally divide $\gamma$ in the middle thus $\mathbb{A}\equiv 0$;
for front-back asymmetric shapes, we divide $\gamma$ at the arclength where $x_1$ is the largest along the generating curve $s=\argmax_{s\in\gamma} x_1(s)$, or the average $s$ if $\argmax$ returns more than one $s$.
Positive $\mathbb{A}$ corresponds to shapes whose head part actuates stronger than its tail part, which indicates that the micro-swimmer is likely to be a puller; similarly negative $\mathbb{A}$ indicates that the micro-swimmer is likely to be a pusher. 
}

\begin{figure}
        \centerline{\includegraphics[scale=1]{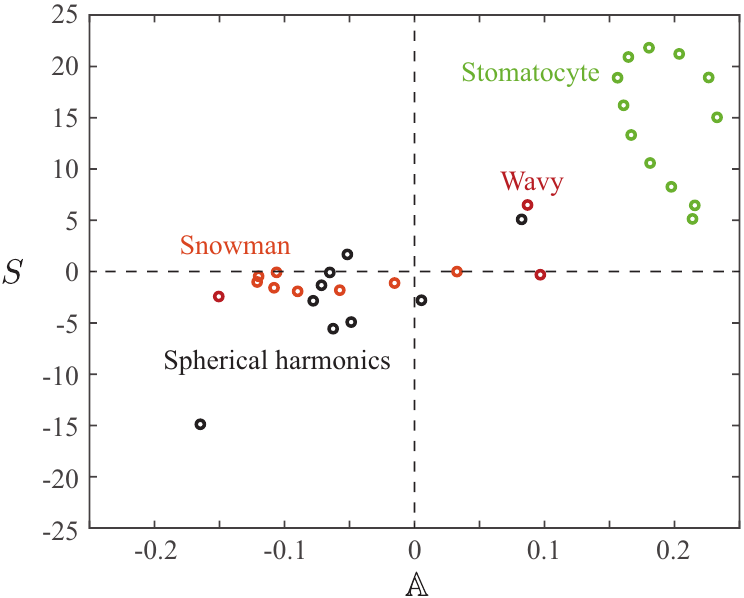}}
	\caption[]{{$\mathbb{A}$ provides a simple prediction of the swimmer type. Swimmers with $\mathbb{A}<0$ are predicted to be pushers ($S<0$), and swimmers with $\mathbb{A}>0$ are predicted to be pullers ($S>0$). Swimmers in the first and third quadrants are correctly predicted. Shape families are shown in Fig.~\ref{fig:gallery} and the generating curves are given in Appendix~\ref{sec:curves}.}}
	\label{fig:prediction}
\end{figure}

{The predictions based on $\mathbb{A}$ for various families of asymmetric shapes are shown in Fig.~\ref{fig:prediction}. Specifically, most of the shapes are correctly predicted as they lie in the first and the third quadrants; the ones that are misclassified, on the other hand, have close-to-zero $\mathbb{A}$ and $S$, which means the head and tail are similarly actuated and the optimal swimmers are close to neutral.
}
%---------------------
%------------------------

\begin{figure}
        \centerline{\includegraphics[width=.9\linewidth]{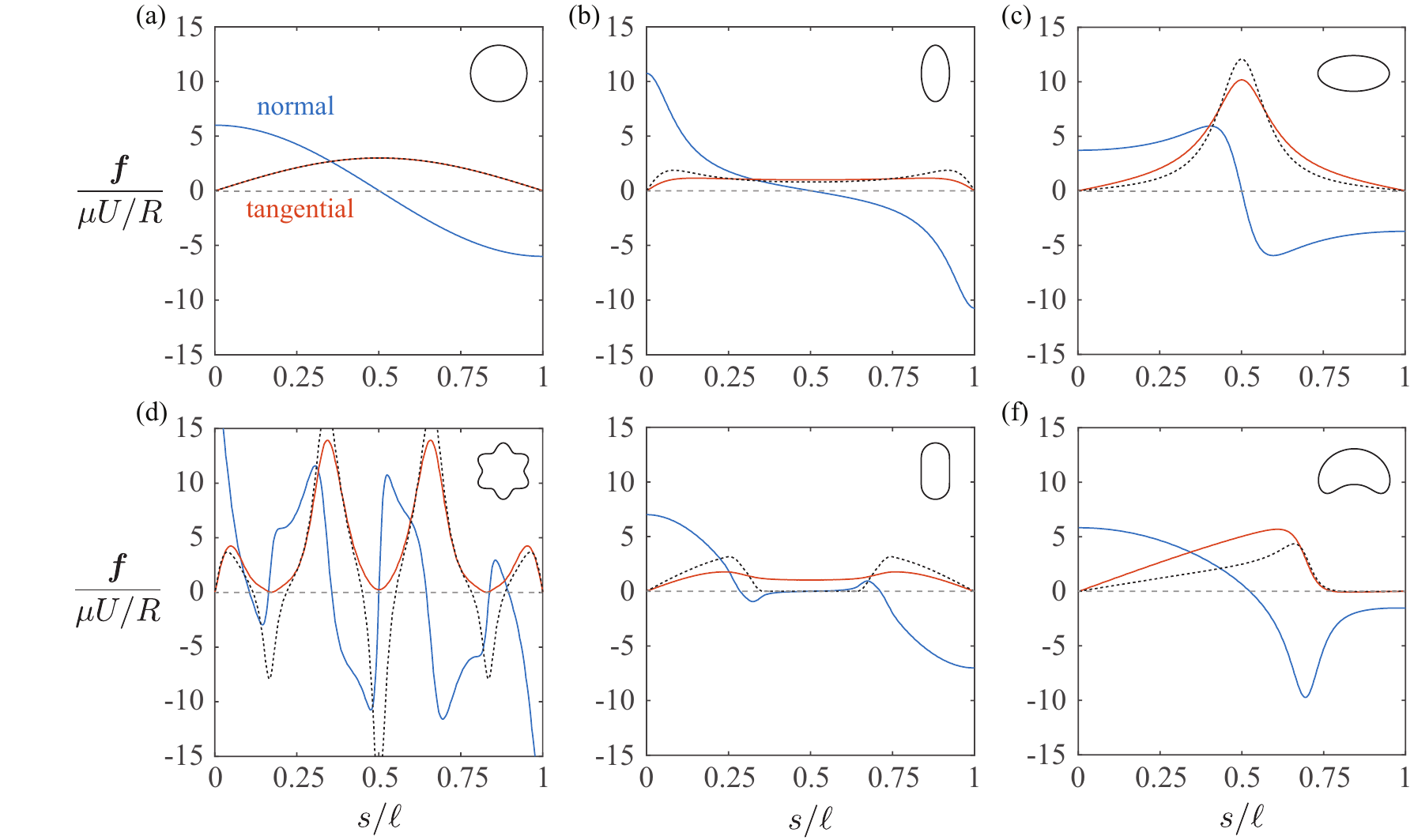}}
	\caption[]{Active force density on the swimmer surface as functions of arc-length along the generating curve. Normal and tangential components of the force densities are depicted by blue and orange curves. Scaled optimal slip velocities $2u^\mathrm{S*}\kappa R/U$ are shown in dotted curves, where $\kappa$ is the local curvature {of the generating curve}. Insets are the shapes of the corresponding swimmers.}
	\label{fig:traction}
\end{figure}

Next, we study the optimal active force density $\boldsymbol{f}$ corresponding to the same shapes. Its normal and tangential components are plotted in Fig.~\ref{fig:traction}. We note that by the no-net-force condition \eqref{eq:nonetforce}, the power loss reduces to $P= 2\pi\int_\gamma \boldsymbol{f}\cdot({u}^{\mathrm{S}}\boldsymbol\tau) x_1\mathrm{d}s,$ implying that only the tangential component contributes to the power loss. 
The change in tangential forces as a function of arclength loosely resembles that of the optimal slip velocity, mediated by the local curvature {of} the generating curve. Qualitatively, a low local curvature suppresses the traction {relative to the slip velocity,} and a high local curvature amplifies it.
Slip velocities scaled by their local curvatures are shown in black dotted curves for a reference.

%---------------------
%------------------------

\begin{figure}
        \centerline{\includegraphics[width=.9\linewidth]{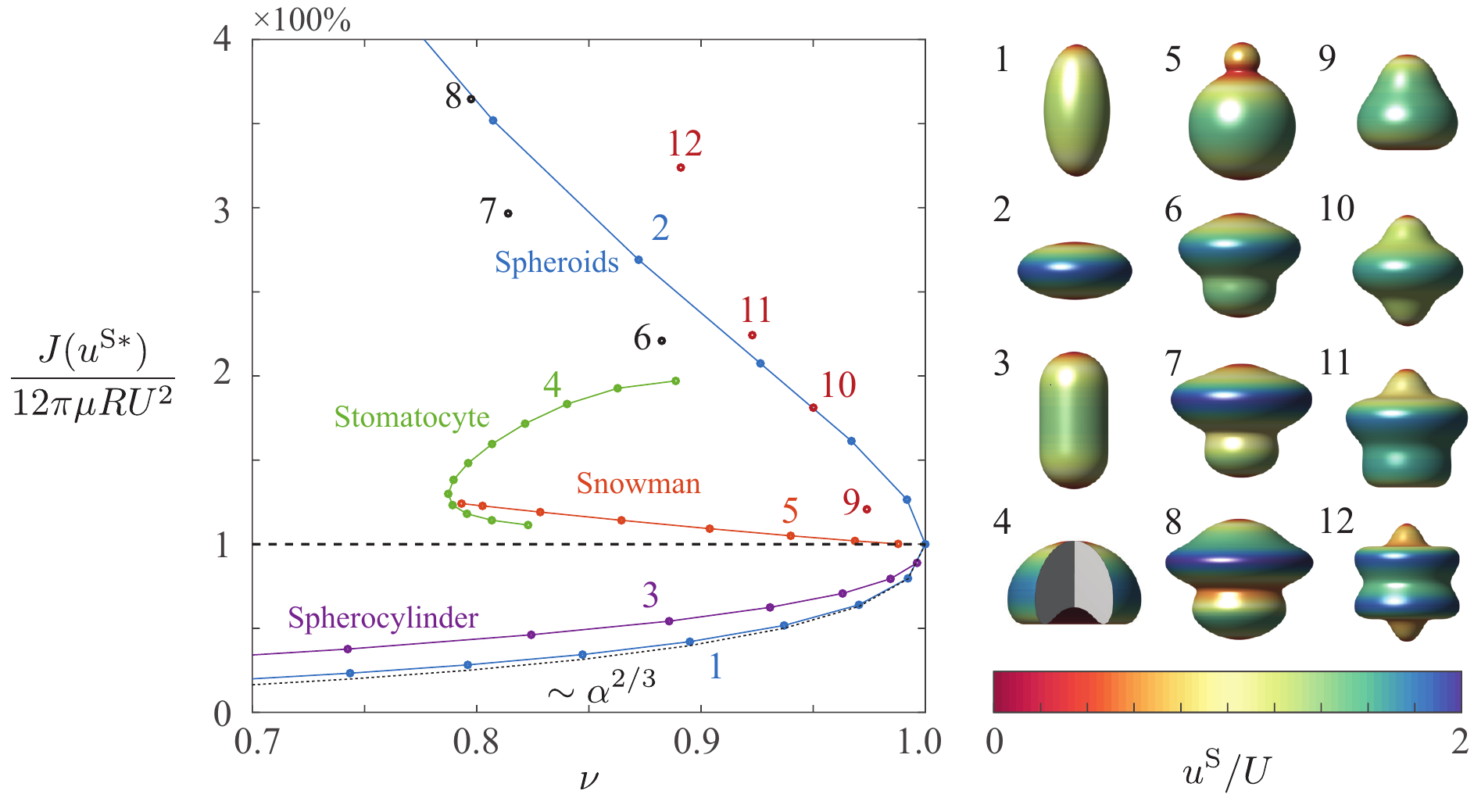}}
	\caption[]{Scaled minimal power loss of different shape families, plotted against the reduced volume $\nu$.  Example shapes are color-coded by the optimal slip velocity. The dotted line shows the {approximation} of power loss given by the slender body theory {$P\sim \mu \alpha^{2/3}U^2$}\citep{leshansky2007frictionless}.}
	\label{fig:gallery}
\end{figure}

%
{In Fig.~\ref{fig:gallery}, we plot the minimal power loss as a function of the reduced volume for various shape families. The power loss is scaled by the minimal power loss of a spherical swimmer with the same volume $J_o = 12\pi \mu RU^2$ with $R = (3V/4\pi)^{1/3}$.}
The minimal power loss for prolate {spheroids} monotonically decreases as the shape gets more slender; in contrast, it is well-known that the shape with the minimal fluid drag is one with approximately 2:1 aspect ratio~\citep{pironneau1973optimum}. By slender body theory, the power loss of a prolate {spheroids} scales as $\sim \mu \alpha^{2/3} U^2$, where $\alpha$ is the aspect ratio~(see \citet{leshansky2007frictionless}). On the other hand, the minimal power loss for oblate {spheroids} grows rapidly as the reduced volume is increased. 
Shapes of the spherocylinder family behave similarly to the prolate {spheroids}, and converge to the spherical case when the length of the cylinder reduces to 0, as expected. It is however worth pointing out that spherocylinder costs more power loss than prolate {spheroids} with the same reduced volume; this relates to the fact that the peak slip velocity for spherocylinder is higher than that of the prolate {spheroids} (Fig.~\ref{fig:example}\,(b)\&(e)).
The stomatocyte family is constructed by `pulling' the rim of the shape, effectively making the shape `taller' and curls deeper and deeper inside. We find that `taller' shapes {require} lower power loss for this shape family, which is qualitatively consistent with the {spheroid} family.  
Finally, we note that the power loss of the snowman family (two spheres attaching with each other) is quite robust to the relative sizes of the two spheres. The power loss is only about $25\%$ higher than that of a single sphere in the limit case where the two spheres are of the same size.

A few other examples that take more generic shapes are also shown in Fig.~\ref{fig:gallery}. The optimal slip velocities are colored on their surfaces while their power loss is shown in the form of scatter points. The generating {curves} of these shapes are formed by spherical harmonics. 
We note that the optimal performance of shapes that appear similar can be very different. For example, the difference in power loss between examples 6 and 8 is about $150\%$ of the spherical swimmer, or $60\%$ of example 6. This result is a strong indicator that the slip velocity of the artificial swimmer, as well as its shape, must be carefully designed to achieve good performance.

{We note that the minimal power loss for all the shape {families} considered here are bounded from below by the curve for prolate {spheroids}. However, since the current work does not optimize shape, whether the prolate spheroids are universally optimal remains to be tested.}

%---------------------
%------------------------

\section{Conclusions}
\label{sc:future}
In this work, we provided a solution procedure for the PDE-constrained optimization problem of finding the optimal slip profile on an axisymmetric micro-swimmer that minimizes the power loss required to maintain a target swimming speed.  While it can be extended to other objective functions, we exploited the quadratic nature of the power loss functional in the control parameters to simplify and streamline the solution procedure.
In the general case, an adjoint formulation and iterative optimization algorithms
can be employed. Regardless of the formulation, however, the use of boundary integral method to solve the Stokes equations greatly reduces the computational cost due to dimensionality reduction. Solving any of the examples presented in this work, for example, required only a few seconds on a standard laptop. Extending our procedure to fully three-dimensional (non-axisymmetric) shapes is straightforward; the key technical challenge is incorporating a high-order boundary integral solver, for which open-source codes are now available (e.g., see~\citet{gimbutas2013fast}).

{Based on our numerical results, we came up with a heuristic metric that can  classify the optimal swimming pattern for a given shape. It measures relative actuation of the `head' and the `tail' of the swimmer and predicts whether the optimal swimmer is head-actuated (puller) or tail-actuated (pusher). This metric could inform the early design of optimal slip for a given shape without the need for carrying out numerical optimization.}

The optimization procedure developed in this work can directly be employed in the design pipeline of autophoretic particles. For example, in the case of diffusiophoresis, the computed optimal slip profile for a given shape can be used to formulate the chemical coating pattern of the phoretic particles. {We acknowledge that the cost function for such optimization may need to be modified accordingly to reflect the chemical nature of the problem~\citep{Sabass2012Dynamics}.} Another natural extension of this work is to relax the steady slip assumption and consider time-periodic squirming motion as done in \cite{Michelin2010Efficiency}. This would be particularly useful for studying the ciliary locomotion of micro-organisms with arbitrary shapes. Furthermore, building on the recent work of \cite{bonnet2020shape}, we are developing solvers for the shape optimization problem of finding the most efficient micro-swimmer shapes under specified area, volume or other physical constraints. 
\\
\section*{Acknowledgement}
Authors gratefully acknowledge support from NSF under grants DMS-1719834 and DMS-1454010. 
{Authors appreciate the constructive suggestions provided by the anonymous referees, which helped them to improve the paper. }

%\section{Appendix}
\appendix
\counterwithin{figure}{section}
\textcolor{black}{
\section{Parameter space }\label{sec:parameterspace}
We parametrize the slip velocity using a piecewise B-spline approximation. 
The slip velocity $u^{\mathrm{S}}(t)$ is determined by $(M+1)$ {\em control points}, $u^{\mathrm{S}}(t_i) = \varphi_i$ for $i = 0,\cdots,M$, and is interpolated by B-spline basis functions between the control points. Here $t\in[0,\pi]$ is a reparametrization of the arc-length $s$.
In theory, we only need to assign control points for $t_i$ between $0$ and $\pi$ to generate an admissible slip velocity by symmetry.
In practice, however, we assign control points in the full period $t_i\in[0,2\pi]$ and impose periodic boundary conditions to determine the spline coefficients, as detailed below.}

\textcolor{black}{
\indent Let $M=2N+2$, where $N$ is the number of {\em free} control points between $0$ and $\pi$.
Let all control points be equally spaced, we have $t_i = 2\pi i/M$, $i = 0,\cdots,M$.
To make sure the slip velocity is axisymmetric, we assign {\em ghost} control points $\varphi_i= -\varphi_{M-i}$ for $N+1<i<2N+2$ and enforce zero conditions at the poles $\varphi_i = 0$, for $i = 0,N+1,2N+2$.
}

\textcolor{black}{
The general B-spline formulation of order 5 is given by
\begin{equation}\label{eq:Bspline}
u^{\mathrm{S}}(t):= \sum_{k=-5}^{M-1} \xi_k B_k(t), \quad t\in[0,2\pi],
\end{equation}
where $B_k(t) = B^*_{k,5}(\frac{M}{2\pi}t)$ is a modified $k$-th B-spline basis function,
and $B^*_{k,p}$ is the standard $k$-th B-spline basis function of degree $p$, given by recurrence
\begin{align}
B^*_{k,0}(t) & = \left\{\begin{array}{ll} 1, & k\le t < k+1 \\ 0,& \mbox{otherwise}  \end{array} \right.\\
    B^*_{k,p} (t) &= \frac{t-k}{p}B^*_{k,p-1} (t) + \frac{p+k+1-t}{p} B^*_{k+1,p-1}(t).
\end{align}
In order to obtain the  $(M+5)$ B-spline coefficients $\xi_k$ from the $(M+1)$ control points $\varphi_i$, we need four more equations to close the system. Specifically, we use the periodic boundary conditions of the derivatives
\begin{equation}
\frac{\mathrm{d}^n u^{\mathrm{S}}}{\mathrm{d} t^n}(0) = \frac{\mathrm{d}^n u^{\mathrm{S}}}{\mathrm{d} t^n}(2\pi), \quad n = 1,2,3,4.
\end{equation}
These system of equations uniquely determine the B-spline coefficient $\xi_k$ from the control points $\varphi_i$. The slip velocity $u^\mathrm{S}(t)$ along the generating curve could then be found by substituting $\xi_k$ into \eqref{eq:Bspline}.
}

\begin{figure}
        \centerline{\includegraphics[width = \linewidth]{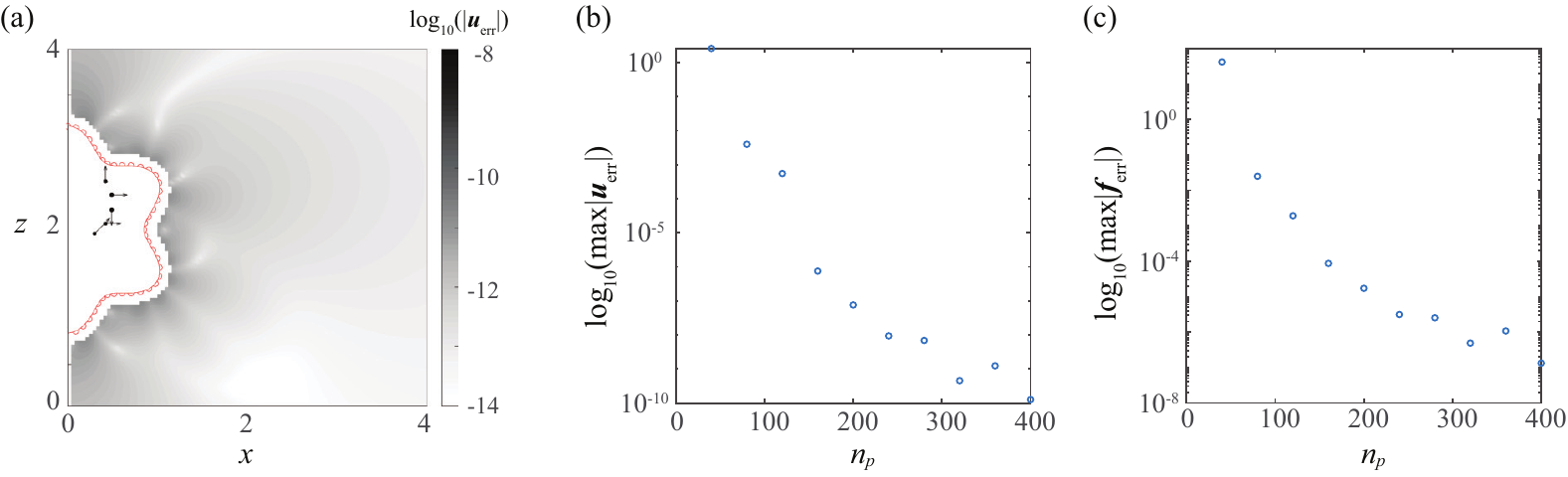}}
\caption[]{(a) The absolute error between the exact solution and the numerical solution with a total of 400 Gaussian quadrature points; color-code represents $\log_{10}(|\boldsymbol{u}_{exa} - \boldsymbol{u}_{num}|)$. (b) The $L_\infty$-norm of the error in the flow field shown as a function of the number of quadrature points. (c) The $L_\infty$-norm of the traction error shown as a function of the number of quadrature points.}
	\label{fig:validation}
\end{figure}

%%%%%%%%%%%%%%%%%%%%
\textcolor{black}{
\section{Numerical validation}\label{sec:validation}
The Green's function $G$ and the traction kernel $T$ used in the ansatz (\ref{eq:BI}) are defined by
\begin{equation}
    \begin{aligned}
        G\left(\boldsymbol{x}-\boldsymbol{y}\right) = \frac{1}{8\pi}\left(\frac{1}{|\boldsymbol{r}|}\boldsymbol{I}+\frac{\boldsymbol{r}\otimes\boldsymbol{r}}{|\boldsymbol{r}|^3}\right),\ \boldsymbol{r}=\boldsymbol{x}-\boldsymbol{y},
    \end{aligned}
\end{equation}
\begin{equation}
    \begin{aligned}
        \boldsymbol{n}\left(\boldsymbol{x}\right) T\left(\boldsymbol{x}-\boldsymbol{y}\right) = -\frac{3}{4\pi}\frac{\boldsymbol{r}\otimes\boldsymbol{r}}{|\boldsymbol{r}|^5}\boldsymbol{r}\cdot\boldsymbol{n}\left(\boldsymbol{x}\right).%\ %\boldsymbol{r}=\boldsymbol{x}-\boldsymbol{y}.
    \end{aligned}
\end{equation}
Due to the rotational symmetry of $\Gamma$, 
we can transform the layer potentials (\ref{eq:BI}) into convultions on the generating curve $\gamma$ by integrating analytically in the $\theta$-direction. The integral kernels take the following form (\citet{veerapaneni2009numerical}): 
\begin{equation}
    \begin{aligned}
         G_\gamma(\boldsymbol{x},\boldsymbol{y}) &=\frac{1}{8\pi}\int_0^{2\pi}  \
              \begin{bmatrix}\frac{\cos \theta}{|\boldsymbol{r}|}+ \frac{\left(y_1\cos \theta-x_1\right)\left(y_1-x_1\cos \theta\right)}{|\boldsymbol{r}|^3} & \frac{\left(y_1\cos \theta-x_1\right)\left(y_3-x_3\right)}{|\boldsymbol{r}|^3}\\ \frac{\left(y_1-x_1\cos v\right)\left(y_3-x_3\right)}{|\boldsymbol{r}|^3} & \frac{1}{|\boldsymbol{r}|}+\frac{\left(y_3-x_3\right)^2}{|\boldsymbol{r}|^3}\end{bmatrix}\, d\theta, \\
         \boldsymbol{n}\left(\boldsymbol{x}\right) T_\gamma(\boldsymbol{x},\boldsymbol{y}) &=-\frac{3}{4\pi}\int_0^{2\pi} \
              \begin{bmatrix} \frac{\left(y_1\cos \theta-x_1\right)\left(y_1-x_1\cos \theta\right)}{|\boldsymbol{r}|^5} & \frac{\left(y_1\cos \theta-x_1\right)\left(y_3-x_3\right)}{|\boldsymbol{r}|^5}\\ \frac{\left(y_1-x_1\cos \theta\right)\left(y_3-x_3\right)}{|\boldsymbol{r}|^5} & \frac{\left(y_3-x_3\right)^2}{|\boldsymbol{r}|^5}\end{bmatrix}\\
            &\hspace{0.9in} \left(n_1\left(y_1\cos \theta-x_1\right)+n_3\left(y_3-x_3\right)\right)\, d\theta.
    \end{aligned}
\label{eq:modalG}
\end{equation}
The velocity and traction can therefore be transformed as: $\boldsymbol{u}\left(\boldsymbol{x}\right)=\int_{\gamma}G_\gamma\left(\boldsymbol{x},\boldsymbol{y}\right)\boldsymbol{\mu}\left(\boldsymbol{y}\right)y_1\,ds$,  $\boldsymbol{f}\left(\boldsymbol{x}\right)=-\frac{1}{2}\boldsymbol{\mu}\left(\boldsymbol{x}\right)+\boldsymbol{n}\left(\boldsymbol{x}\right)\int_{\gamma}T_\gamma\left(\boldsymbol{x},\boldsymbol{y}\right)\boldsymbol{\mu}\left(\boldsymbol{y}\right)y_1\,ds$. {The analytic solution of the integrals (\ref{eq:modalG}) can be found in \citet{veerapaneni2009numerical} and \citet[Page 40]{Pozrikidis1992}}.\\
\indent To validate our boundary integral method, we construct a boundary value problem and test the algorithm against the exact solution. As is standard practice, we consider the flow field generated by a set of axisymmetric Stokeslets and the corresponding traction:
\begin{equation}\label{eq:vel_ref}
\boldsymbol{u}_{exa}(\boldsymbol{x}) =  \sum_{k=1}^N G_\gamma(\boldsymbol{x},\boldsymbol{y}_k){\boldsymbol{\tau}_{k}}y_{k,1}, \quad
\boldsymbol{f}_{exa}(\gamma) =  \boldsymbol{n}\left(\gamma\right)\sum_{k=1}^N T_\gamma(\gamma,\boldsymbol{y}_k){\boldsymbol{\tau}_k}(k)y_{k,1},
\end{equation}
where $\{\boldsymbol{y}_k\}$ and $\{\boldsymbol{\tau}_k\}$ are the location and strength of the $k$-th Stokeslet. We randomly choose $5$ Stokeslets whose locations and strengths are given in Fig.~\ref{fig:validation}(a) by the black arrows and substitute them into \eqref{eq:vel_ref} as our reference case.
}

\begin{figure}
        \centerline{\includegraphics[scale = 1]{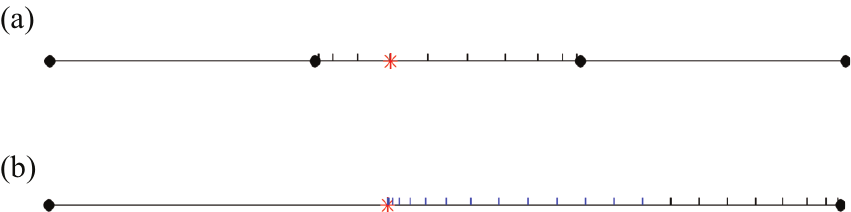}}
\caption[]{(a) Example of a panel with $10$-point Gaussian nodes, and its neighbor panels. The red asterisk is the target. (b) Three panels in (a) are combined into one big panel. The big panel is further divided into two panels by the desired target. Blue grid is a $16$th-order Alpert quadrature rule. And black grid is an $8$-point smooth quadrature rule.}
	\label{fig:panelconfig}
\end{figure}

\textcolor{black}{
To obtain the numerical solution, we first evaluate the reference flow field on the generating curve $\boldsymbol{u}_{exa}(\gamma)$, then treat $\boldsymbol{u}_{exa}(\gamma)$ as the boundary condition to obtain the density vector $\boldsymbol\mu$. The generating curve $\gamma$ is discretized into non-overlapping panels $\gamma = \sum_{p=1}^{N_p}\Lambda_p$. Then on each panel, we place the nodes of a $10$-point Gaussian quadrature. The integral operator can then be approximated by the standard Nystr\"om matrix at these collocation points. The logarithmic singularity is resolved with Alpert quadrature using node locations off the Gauss-Legendre grid \citep{hao2014high}, as illustrated in Fig.~\ref{fig:panelconfig}(a)\,\&(b). Integral of  $G_\gamma\left(\boldsymbol{x},\boldsymbol{y}\right)$ and $T_\gamma\left(\boldsymbol{x},\boldsymbol{y}\right)$ at the desired target, endpoints of two panels in Fig.~\ref{fig:panelconfig}(b), are approximated using correction nodes. Note that two end panels need to be further split adaptively corresponding to north and south poles, until the first and last Gaussian nodes have adjacent neighbors. 
We subsequently use the density vector $\boldsymbol\mu$ to evaluate the numerical solution $\boldsymbol{u}_{num}(\boldsymbol{x})$ outside the microswimmer's surface.
The traction on the generating curve is evaluated from the same density vector $\boldsymbol\mu$ using the traction kernel $\boldsymbol{f}_{num}(\gamma)=-\frac{1}{2}\boldsymbol{\mu}(\gamma)+ \boldsymbol{n}\left(\gamma\right)\int_{\gamma} T_\gamma(\gamma, \boldsymbol{y})\boldsymbol{\mu}\left(\boldsymbol{y}\right)y_1\,ds$.
}

\textcolor{black}{
The absolute error of the numerical solution $\boldsymbol{u}_{num}$ for this example is shown in Fig.~\ref{fig:validation}(a). As can be observed from Fig.~\ref{fig:validation}(b)\,\&(c), our forward solver achieves 10-digit accuracy in the flow field and 6-digit accuracy for traction with 400 quadrature points on the generating curve. For all the test cases presented in Section~\ref{}, 600 Gauss-Legendre quadrature points were used.
}

\begin{figure}
        \centerline{\includegraphics[scale = 1]{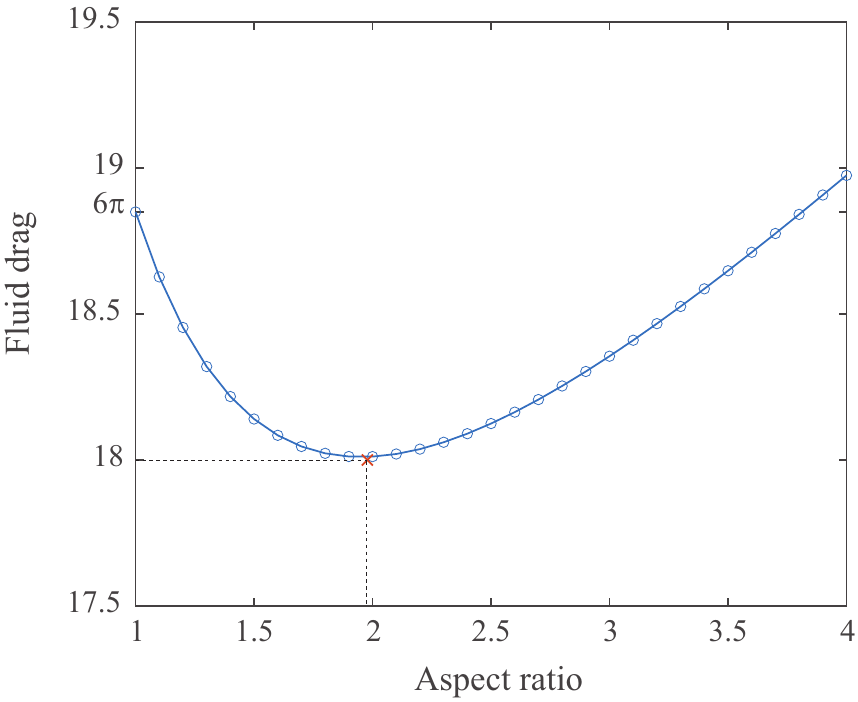}}
\caption[]{Fluid drag of towing a prolate spheroid with unit speed. All spheroids are of the same volume as the unit sphere. The red cross denotes the fluid drag of the optimal profile that minimizes the fluid drag given by \citet{pironneau1973optimum}.}
	\label{fig:validation2}
\end{figure}

\textcolor{black}{
As a further validation of our numerical scheme, we computed the fluid drag of a family of prolate and oblate ellipsoids. The shape that yields the minimal fluid drag is a prolate ellipsoid with a roughly $2:1$ aspect ratio (Fig.~\ref{fig:validation2}), consistent with the optimal shape obtained previously in \citet{pironneau1973optimum}.
}

%---------------------
%------------------------

%---------------------
%------------------------

\textcolor{black}{
\section{Generating curves of the shapes used in the paper}\label{sec:curves}
Here, for reproducibility purposes, we list equations of all the generating curves used in this paper. 
In all cases below, $i=\sqrt{-1}$, $t\in[0,\pi]$ is the polar angle, the equations are defined on the complex plane and the axis of symmetry is the imaginary axis.
\begin{itemize}
\item \hspace{.03in} {\em Spheroids:} $z = \alpha^{-1/3} \sin(t) + i \alpha^{2/3}\cos(t)$, $\alpha$ is the aspect ratio. 
\item \hspace{.03in} {\em Wavy shapes:} $z = (1+0.15\cos(kt) \exp(i (\pi/2-t)))$, $k\in\{3,4,5,6\}$ is the order of the perturbation.
\item \hspace{.03in} {\em Stomatocyte:} $z = (1.5+\cos t)(\sin(\lambda \pi\sin t) + i\cos(\lambda\pi\sin t)) - 0.5i$, $\lambda\in[0.4, 0.95]$ controls the vertical `stretchiness' of the shape.
\item \hspace{.03in} {\em Harmonics:} $z = \rho(t) \sin t - i\rho(t) \cos t$, where $\rho(t) = 1+ r Y_n^m(t,0)$, where $Y_n^m(\theta, \varphi)$ is the spherical harmonics of degree $n$ and order $m$, evaluated at the colatitude $\theta$ and longitude $\varphi$. %Different values of $m, n, r$ are used to generate shapes \#6, 7, 8 in the gallery. 6: (m,n,r) = (7,8,0.5), 7: (m,n,r) = (7,8,rand?), 8: (m,n,r) = (15,16,0.5),
\item \hspace{.03in} {\em Spherocylinder} shapes were generated by simply attaching semi-spherical caps to a cylinder with the same radius and subsequently smoothing using B-splines upto order 5. 
\item \hspace{.03in} {\em Snowman} shapes were generated by two spheres of different radii glued together with the centroid distance set to $90\%$ of the sum of the radii, followed by smoothing.
\end{itemize}
}
\bibliographystyle{jfm}
%\bibliography{refs}
\bibliography{references}

\end{document}